\documentclass[aps,jcp,twocolumn,groupedaddress,longbibliography]{revtex4-1}
\usepackage{graphicx,amsmath,amssymb,color}
\begin{document}

\title{Factors influencing thermal solidification of bent-core trimers}
\author{Elvin D. Salcedo$^1$}
\author{Hong T. Nguyen$^2$}
\author{Robert S. Hoy$^1$}
\email{rshoy@usf.edu}
\affiliation{$^1$Department of Physics, University of South Florida, Tampa, FL 33620}
\affiliation{$^2$Department of Materials Science and Engineering, University of Texas at Dallas, Richardson, Texas 75080}
\date{\today}
\begin{abstract}
Bent-core trimers are a simple model system for which the competition between crystallization and glass-formation can be tuned by varying a single parameter:\ the bond angle $\theta_0$.  
Using molecular dynamics simulations, we examine how varying $\theta_0$ affects their thermal solidification.  
By examining trends with $\theta_0$, comparing these to trends in trimers' jamming phenomenology, and then focusing on the six $\theta_0$ that are commensurable with close-packed crystalline order, we obtain three key results: \textbf{(i)} the increase in trimers' solidification temperature $T_s(\theta_0)$ as they straighten (as $\theta_0 \to 0^\circ$) is driven by the same gradual loss of \textit{effective} configurational freedom that drives athermal trimers' decreasing $\phi_J(\theta_0)$; \textbf{(ii)} $\theta_0$ that allow formation of both FCC and HCP order crystallize, while $\theta_0$ that only allow formation of HCP order glass-form; \textbf{(iii)}  local cluster-level structure at temperatures slightly \textit{above} $T_s(\theta_0)$ is highly predictive of whether trimers will crystallize or glass-form.
\end{abstract}
\maketitle

\section{Introduction}
\label{sec:intro}

\begin{figure}
\includegraphics[width=3in]{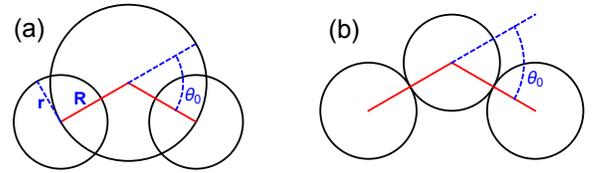}
\caption{Rigid bent-core trimers with bond angle $\theta_0$.  Panel (a) shows the general geometry with unspecified ($r,\ R$).  Here we study the $r = R = 1$ case shown in panel (b).}\label{fig:trimermodel}
\end{figure}

Bent-core trimers are a simple model for many small organic molecules whose bulk liquids exhibit both crystallization and glass-formation in experiments \cite{andrews55,alba90,powell14,ping11}.
Their shape can be characterized using three parameters (Fig.\ \ref{fig:trimermodel}): the bond angle $\theta_0$, the ratio $r$ of end-monomer radius to center-monomer radius, and the ratio $R$ of intermonomer bond length to center-monomer diameter.
It is well known that the properties of systems composed of such molecules depend strongly on all three of these parameters.
For example, the three terphenyl isomers, which can be modeled as trimers with the same $r$ and $R$ but different $\theta_0$, form very differently structured bulk solids 
under the same preparation protocol \cite{andrews55}. 
Analogous $\theta_0$-dependent differences occur between the three xylene isomers \cite{alba90}.
Two more classes of small molecules, the diphenylcycloalkenes and cyclic stilbenes, which can each be modeled as having the same $\theta_0$ but different $r$ and $R$, show similarly complex and poorly-understood dependence of crystallizability on molecular shape  \cite{powell14,ping11}.

One of the reasons why our understanding of such phenomena and hence our ability to engineer crystallizability/glass-formability at the molecular level remains very limited is that only a few theoretical studies have isolated the role played by molecular shape using simple models.
Ref.\ \cite{jennings15} reported the densest packings of 2D $R = 1/2$ trimers as a function of $r$ and $\theta_0$.
Refs.\  \cite{pedersen11,pedersen11b,pedersen19} examined the thermal solidification of Lewis-Wahnstrom-like models ($R = 2^{-1/6},\ r = 1,\ \theta_0 = 105^\circ$ \cite{lewis94}) and reported several nontrivial effects of trimeric structure, e.g.\ that its enhancement of the interfacial energy between crystalline and liquid phases promotes glass-formation.
The tangent-sphere ($r = R = 1$) case shown in Fig.\ \ref{fig:trimermodel}(b) is of considerable interest because it allows straightforward connection to results obtained for monomers -- and hence isolation of the role played by the bond and angular constraints -- while remaining a reasonable minimal model for trimeric molecules.
Refs.\ \cite{griffith18,griffith19} reported the densest packings of 2D and 3D tangent-sphere trimers and contrasted them to the jammed packings they form under dynamic athermal compression.

In this paper, we examine the thermal solidification of bent-core tangent-sphere trimers as a function of their bond angle $\theta_0$ using molecular dynamics simulations.
We obtain three key results: \textbf{(i)} the sharply increasing solidification temperature $T_s(\theta_0)$ for $\theta_0 \lesssim 20^\circ$ is driven by the same gradual loss of \textit{effective} configurational freedom that drives athermal trimers' decreasing $\phi_J(\theta_0)$ \cite{griffith19}; \textbf{(ii)} $\theta_0$ that allow formation of both FCC and HCP order crystallize, while $\theta_0$ that only allow formation of HCP order glass-form; \textbf{(iii)} measurements of local cluster-level structure via CCE and TCC analyses \cite{cce09,malins13d} are as predictive of whether trimers will crystallize or glass-form as they are for monomeric systems.

\section{Model and Methods}
\label{sec:methods}

\subsection{Molecular dynamics simulations}
\label{subsec:MD}

Each simulated trimer contains three monomers of mass $m$.  
These trimers are rigid; their bond lengths and angles are held fixed by holonomic constraints.
Monomers belonging to different trimers interact via the truncated and shifted Lennard-Jones potential
\begin{equation}
U_{LJ}(r) = \epsilon\left[\left(\displaystyle\frac{\sigma}{r}\right)^{12} - \left(\displaystyle\frac{\sigma}{r_c}\right)^{12} - 2\left(\left(\displaystyle\frac{\sigma}{r}\right)^{6} - \left(\displaystyle\frac{\sigma}{r_c}\right)^{6}\right)\right],
\label{eq:LJpot}
\end{equation}
where $\epsilon$ is the energy scale of the pair interactions, $\sigma$ is monomer diameter, and $r_c = 2^{7/6}\sigma$ is the cutoff radius.

The Lewis-Wahnstrom model \cite{lewis94} is a good glass-former largely because its equilibrium backbone bond length $\ell_0$ is incommensurable with its equilibrium nearest neighbor distance for nonbonded neighbors $r_0$; specifically, it has $\ell_0 = \sigma$ and $r_0 = 2^{1/6}\sigma$. 
In contrast, the current model makes these lengths commensurable ($\ell_0 = r_0 = \sigma$).
Any frustration against crystallization is driven primarily by the bond angle $\theta_0$ \cite{otherfrust}.
Six $\theta_0$ are commensurable with 3D close-packing:\ $0^\circ,\  \cos^{-1}(5/6)  \simeq 33.5^\circ,\ 60^\circ,\ \cos^{-1}(1/3) \simeq 70.5^\circ,\ 90^\circ,\ \rm{and}\ 120^\circ$ \cite{griffith19}.  
All of these $\theta_0$ allow formation of HCP crystals, whereas only $0^\circ,\ 60^\circ,\ 90^\circ,\ \rm{and}\ 120^\circ$ allow formation of FCC crystals.
We will argue below that this distinction is critical.

Initial states are generated by placing $n_{tri} = 1333$ trimers randomly within a cubic cell at a packing fraction $\phi_0 = .5$.
Periodic boundary conditions are applied along all three directions and Newton's equations of motion are integrated with a timestep $\delta t = .005\tau$, where the unit of time is $\tau=\sqrt{m\sigma^2/\epsilon}$.
Systems are equilibrated at $k_B T = \epsilon$ until intertrimer structure has converged, then slowly cooled to $T=0$ [at a rate $\dot{T} = 10^{-6}\epsilon/(k_B \tau)$], all at zero pressure.
Fixed covalent bond lengths and bond angles are maintained using a standard method \cite{kamberaj05}.
All MD simulations are performed using LAMMPS \cite{plimpton95}.

We identify the solidification temperature $T_s$ of each system using one of two methods.
For systems exhibiting crystallization, we simply locate the center of the first-order-like jump in $\phi(T)$ as $T$ decreases.
For systems exhibiting glass-formation, we identify $T_s$ as the intersection points of linear fits to the high- and low-$T$ portions of their $\phi(T)$ curves.
Consistent with our focus on trimers' solidification under \textit{dynamic} cooling, these $T_s(\theta_0)$ are below the equilibrium solid-liquid transition temperatures $T_{\rm melt}(\theta_0)$: $T_s(\theta_0) = T_{\rm melt}(\theta_0) -  \Delta T(\theta_0,\dot{T})$, where  $\Delta T$ increases with $\dot{T}$ and with glass-formability.
The $\dot{T}$ employed here is low enough that $\Delta T(\theta_0)$ is very small for our best crystal-formers.

\subsection{Measures of local order}
\label{subsec:CCETCC}

We characterize systems' local structure using the characteristic crystallographic element (CCE) \cite{cce09} and topological cluster classification (TCC)  \cite{malins13d} methods.
Recent studies employing CCE \cite{karayiannis11,karayiannis12,karayiannis13} or TCC \cite{taffs10,taffs13,malins13a,malins13b,royall15} analyses have led to much progress in our understanding of thermal solidification.
These studies have found that the presence of energetically stable amorphous clusters in a system's liquid state (at temperatures slightly above its $T_{\rm melt}$) strongly promotes glass-formation.
Long-lived clusters that are fivefold-symmetric and/or are subsets of icosahedra are particularly effective glass-promoters \cite{karayiannis11,karayiannis12,malins13a,malins13b}.

CCE employs descriptors known as ``norms'' that quantify the orientational and radial similarities of a given monomer's local environment to that of various reference structures such as HCP and FCC crystals \cite{cce09}.
These norms are built around sets of point symmetry groups that uniquely characterize the reference structure. 
Criteria based on these norms are applied to determine whether the monomer can be associated with structure $\emph{X}$.
Here we identify monomers as FCC-like, HCP-like, or fivefold-like if their respective norms $n_{\rm X}$ (where $X\ =\ \rm{FCC},\ \rm{HCP},\ \rm{or\ 5f}$) are less than $0.21$.  
Monomers that satisfy none of these conditions are classified as ``other''; these typically possess locally amorphous order.
The relevant mathematical formulae and the algorithmic implementation are described in Ref.\ \cite{cce09}.

\begin{figure}[h]
\includegraphics[width=3.375in]{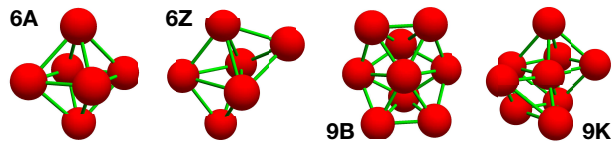}
\caption{The four clusters of primary interest for our TCC analyses.  We will show that the $\theta_0$- and $T$-dependencies of these clusters' formation propensities are strongly coupled.}
\label{fig:clusts}
\end{figure}

TCC is similar in spirit to CCE, but identifies differently structured clusters by their differing bond topology.
Here we employ TCC to track how the populations of various microstructural motifs within our systems vary with $\theta_0$ and $T$, using the same procedures detailed in Ref.\ \cite{malins13d}. 
We focus on the four cluster types shown in Figure \ref{fig:clusts}.
$6A$ is the octahedron, a common motif in close-packed crystals.  
$6Z$, a polytetrahedral structure, has higher energy but also much higher entropy than $6A$ \cite{meng10}, and is a common motif in glassy and jammed systems \cite{anikeenko07}.
$9B$ is a partial icosahedron; recall that icosahedral order has long been known to promote glass-formation \cite{frank52}.
Finally, $9K$ is a subset of the HCP lattice.

During the cooling runs, we monitor the fractions $f_{\rm X}(T)$ of particles identified as X-like (for CCE analyses) or belonging to at least one cluster of type X (for TCC analyses).
We will show below that $\theta_0$-dependent differences in how these quantities evolve with decreasing $T$ predict whether a given system will crystallize or glass-form.

\section{Results}
\label{sec:results}

\subsection{Macroscopic}

\begin{figure}[h]
\includegraphics[width=3.125in]{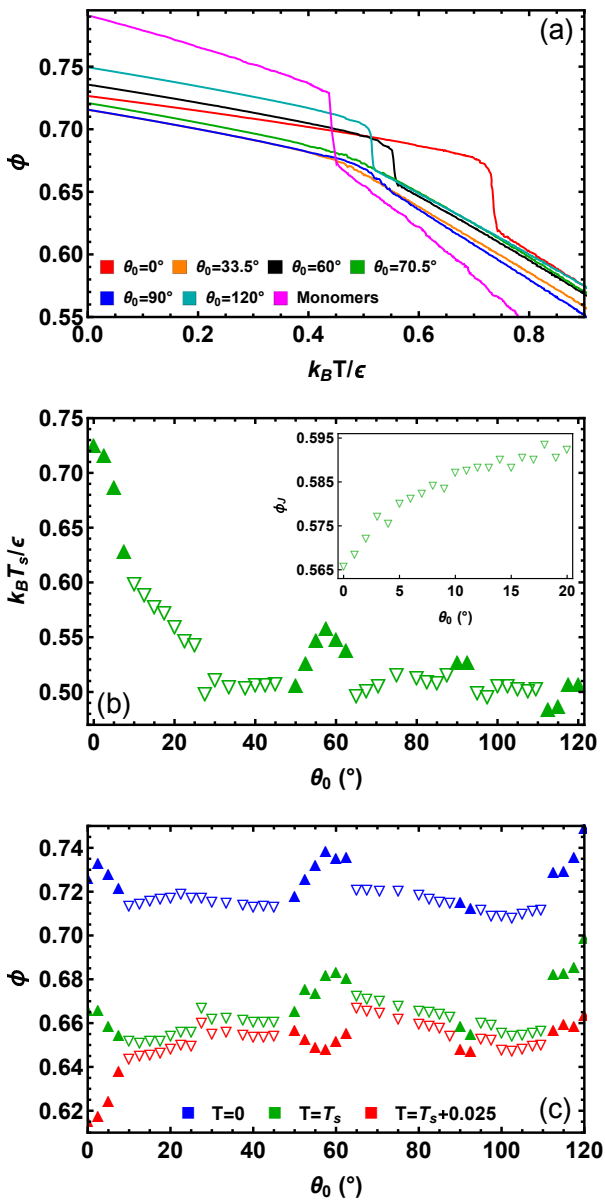}
\caption{Basic solidification results for bent-core trimers. Panel (a):\ Packing fraction $\phi(T)$ 
for six representative $\theta_0$ and for monomers \cite{gtcp}.  Panel (b):\ Solidification temperatures $T_s(\theta_0)$ for all $\theta_0$, calculated as described in Section \ref{subsec:MD}.  The inset compares these to athermal bent-core trimers' $\phi_J(\theta_0)$ for $\theta_0 \leq 20^\circ$ \cite{griffith19}.  Panel (c):\ Packing fractions for all $\theta_0$ at three representative temperatures, where the $T_s(\theta_0)$ values are the same as in panel (b).   In panels (b-c), upward (downward) triangles indicate crystal-formers (glass-formers). }
\label{fig:phiTs}
\end{figure}

Figure \ref{fig:phiTs} summarizes the most basic features of bent-core trimers' response to slow thermal cooling.
Panel (a) shows the packing fraction $\phi(T) = \pi\rho(T)/6$ -- where $\rho = 3n_{tri}/V$ is the monomer number density -- for the six values of $\theta_0$ that are commensurable with 3D close-packing, and contrasts these to results for monomers.
Three features are immediately apparent.
First, some $\theta_0$ produce sharp, first-order-like jumps in $\phi(T)$ that indicate rapid crystallization, while others produce typical glassy behavior where a smooth crossover regime centered at $T \simeq T_g(\theta_0)$ connects roughly linear behavior at high and low $T$.
It is noteworthy that some $\theta_0$ that are commensurable with formation of close-packed crystals glass-form even at the low cooling rate ($\dot{T} = 10^{-6}/\tau$) employed here.
Second, even for those systems that do crystallize, the sharpness of the transition depends strongly on $\theta_0$; systems with sharper transitions (e.g.\ $\theta_0 = 120^\circ$) have faster crystallization kinetics.
Third, trimers always solidify at higher temperatures and lower densities than their monomeric counterparts \cite{footmonsolid}.

As shown in panel (b), trimers' solidification temperatures remain relatively constant as $\theta_0$ decreases from $120^\circ$ to $\sim 20^\circ$, then increase rapidly as $\theta_0 \rightarrow 0^\circ$.
This trend presumably results from the loss of \textit{effective} configurational freedom as trimers approach linearity.
Specifically, the middle monomer in a bent trimer can relax away from obstacles by rotating about the line connecting the end monomers even if the end monomers are held fixed, whereas the middle monomer in a straight trimer cannot.
The same decrease in in effective configurational freedom appears to drive the decreasing $\phi_J$ in athermal bent-core trimers as $\theta_0  \rightarrow 0^\circ$ \cite{griffith19}, and analogous phenomena appear to decrease $\phi_J$ and increase $T_s$ in model polymeric systems \cite{hoy17,plaza17}.

Panel (c) shows the packing fractions $\phi(T)$ for all $\theta_0$ at three representative temperatures: $T = 0$, $T = T_s(\theta_0)$, and $T = T_s(\theta_0) + .025\epsilon/k_{B}$.
The final densities reached at $T = 0$ do not depend very strongly on $\theta_0$; the densest system ($\theta_0 = 120^\circ$) is only $\sim 6\%$ more tightly packed than the least-dense system ($\theta_0 = 102.5^\circ$).
The broad minimum around $\theta_0 = 102.5^\circ$ is consistent with -- and, to some extent \cite{lwcaveat}, supports -- Lewis-Wahnstr{\"o}m-like models' choice of $\theta_0 = 105^\circ$ as a bond angle suitable for modeling glassforming trimer-like molecules such as OTP \cite{lewis94,pedersen11,pedersen11b,pedersen19}. 
Similar trends with $\theta_0$ are apparent at $T = T_s$; the $\theta_0$-dependence of $\phi(0)-\phi(T_s)$ is fairly weak.
In contrast, the much stronger $\theta_0$-dependence of $\phi(T_s)-\phi( T_s + .025\epsilon/k_B)$ reflects the trends shown in panel (b), i.e. small-$\theta_0$ trimers crystallize from much-lower-density supercooled liquids than their large-$\theta_0$ counterparts.

\begin{figure}
\includegraphics[width=3.25in]{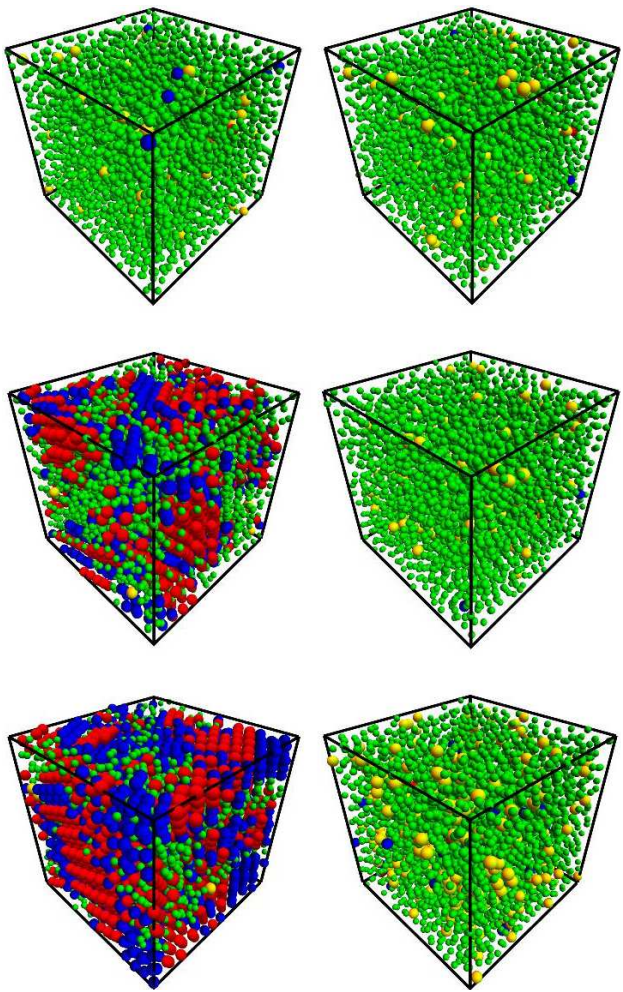}
\caption{Snapshots of the best-crystallizing ($\theta_0 = 120^\circ$; left panels) and best-glassforming ($\theta_0 = 70.5^\circ$; right panels) systems at $T = T_s + .025\epsilon/k_B$ (top panels), $T = T_s$ (middle panels), and $T = 0$ (bottom panels). Red, blue, yellow and green colors respectively indicate monomers classified as FCC, HCP, fivefold-symmetric, and ``other''.}
\label{fig:snaps}
\end{figure}

\subsection{Microscopic}

The wide range of thermal solidification behavior highlighted in Fig.\ \ref{fig:phiTs} results entirely from changing the bond angle $\theta_0$, i.e.\ from changing trimers' shape.
This suggests that it can be understood at a microscopic level by applying tools and concepts like those applied to model atomic/colloidal systems in Refs.\  \cite{karayiannis11,karayiannis12,karayiannis13,taffs10,taffs13,malins13a,malins13b,royall15}.

Figure \ref{fig:snaps} shows snapshots of $\theta_0 = 70.5^\circ$ and  $\theta_0 = 120^\circ$ systems at $T = T_s + .025\epsilon/k_B,\ T_s,\ \rm{and}\ 0$, with monomers color-coded by their CCE norms.
The systems appear similar at $T_s + .025\epsilon/k_B$; while a few fivefold-symmetric sites are present, most lack any distinguishable order.
For the $\theta_0 = 70.5^\circ$ systems, the only obvious changes as $T$ decreases are that more fivefold-symmetric sites appear, as do a very small number of HCP-ordered sites.
The behavior of $\theta_0 = 120^\circ$ trimers is very different.
A large degree of mixed FCC/HCP order is present by $T = T_s$, and a relatively well-ordered crystal develops as $T$ decreases further; the $T = 0$ snapshot clearly shows crystalline grains separated by stacking faults and amorphous interphase.
Since these two $\theta_0$ respectively have the lowest and highest fraction of close-packed monomers at $T=0$, we characterize them as the best glass-former and best crystal-former.

\begin{figure}[h]
\includegraphics[width=3.2in]{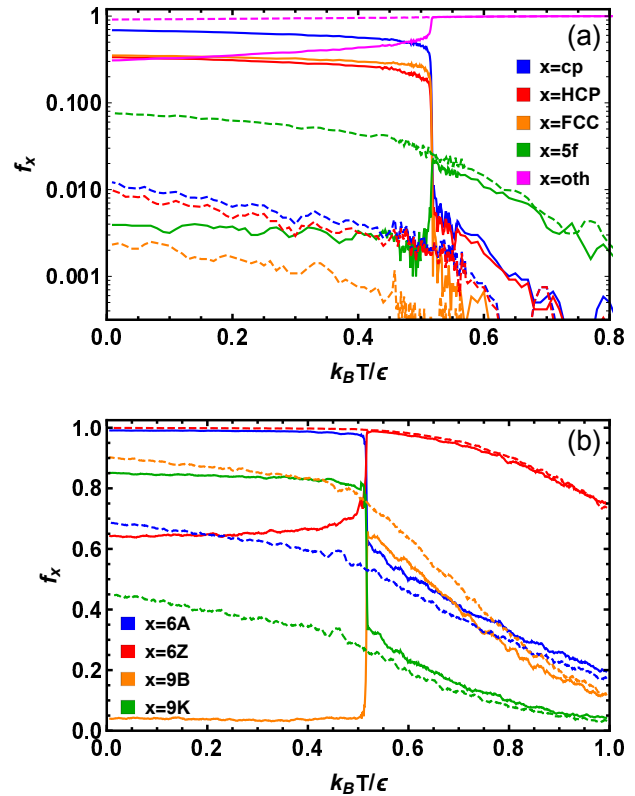}
\caption{CCE and TCC results:\ $f_{\rm X}(T)$.  Solid and dashed curves respectively show results for the best crystal-former ($\theta_0 = 120^\circ$) and the best glassformer ($\theta_0 = 70.5^\circ$).  To suppress finite-system-size noise, data for $|T - T_s| > .02\epsilon/k_B$ are smoothed using moving averages.  The negative $\partial f_{\rm X}/\partial T$ for $T \gtrsim T_s + 0.1\epsilon/k_B$ in panel (b) are driven primarily by changes in $\phi$ (i.e.\ thermal expansion) rather than changes in orientational order; TCC counts monomers as ``bonded'' only if the distance between them is $\leq 1.2\sigma$.}  
\label{fig:TCC1}
\end{figure}

We now examine these differences more quantitatively.
Figure \ref{fig:TCC1} shows $f_{\rm X}(T)$ for these systems.
Panel (a) illustrates how the crystal-former shows sharp (and typical \cite{karayiannis11,karayiannis12}) upward jumps in both $f_{\rm FCC}$ and $f_{\rm HCP}$ and sharp downward jumps in $f_{5f}$ and $f_{oth}$ at $T=T_s$, whereas the glass-former shows no such jumps.
Intriguingly, $f_{cp}$ is about an order of magnitude higher in the crystal-former at temperatures slightly above $T_s$ than it is in the glass-former, presumably indicating the crystal-forming liquid's greater population of subcritical nuclei.
Results from TCC analyses [panel (b)] show analogous trends; the crystal-former shows sharp (and typical \cite{taffs13,royall15}) upward jumps in both $f_{6A}$ and $f_{9K}$ and sharp downward jumps in $f_{6Z}$ and $f_{9B}$ at $T=T_s$, whereas the glass-former does not.
Remarkably, the formation propensity of midsize crystalline clusters, e.g.\ 9K [midsize amorphous clusters, e.g.\ 9B] is is significantly higher [lower]  in the crystal-former even at $T \simeq 1.5T_s$.

These differences are consistent with trends seen in many model colloidal systems  \cite{karayiannis11,karayiannis12,karayiannis13,royall15,taffs10,taffs13,malins13a,malins13b}. 
Considered in isolation, they are not surprising.
What \textit{is} surprising is that they are so stark despite the fact that both   $\theta_0 = 70.5^\circ$ and $\theta_0 = 120^\circ$ trimers can form 3D close-packed crystals \cite{griffith19}.
We now attempt to understand them further by comparing CCE and TCC results for all $\theta_0$.

\begin{figure}
\includegraphics[width=3.2in]{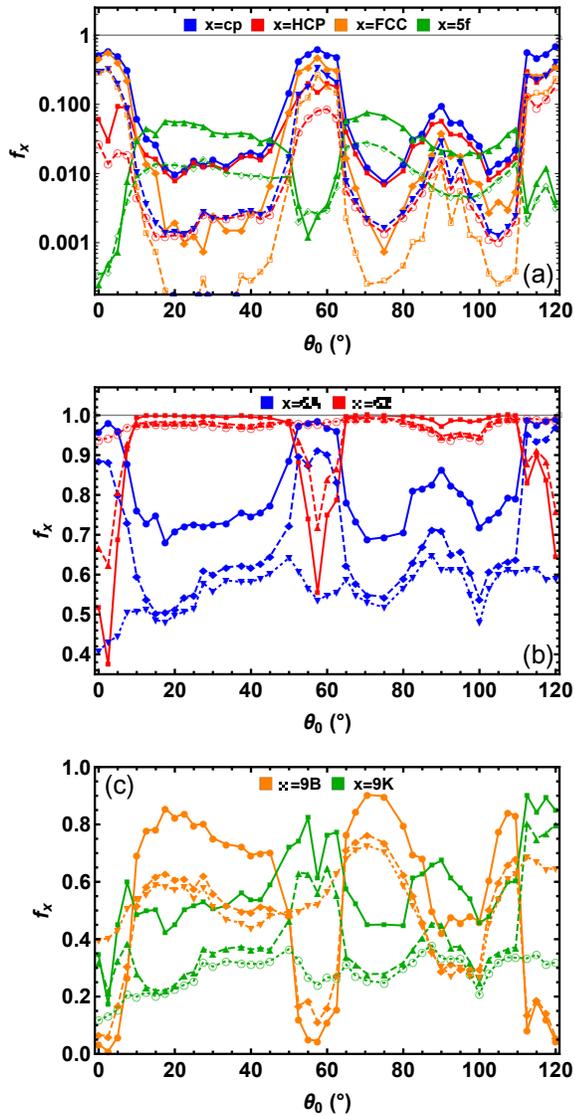}
\caption{CCE and TCC results:\ $f_{\rm X}$ for all $\theta_0$ at selected $T$.  Solid, dashed, and dotted curves respectively show $f_{\rm x}(\theta_0)$ at $T = 0$, $T = T_s(\theta_0)$, and  $T = T_s(\theta_0) + .025\epsilon/k_B$.  Note that incommensurability of $\theta_0$ with 3D close-packing does not prelude formation of small close-packed crystallites \cite{griffith19}.}  
\label{fig:TCC2}
\end{figure}

Figure \ref{fig:TCC2} shows the $\theta_0$-dependence of $f_{\rm X}(T)$ for $T = 0,\ T_s,\ \rm{and}\ T_s + .025\epsilon/k_B$.
Panel (a) summarizes the essential features of the bond angle's effect on crystallizability.
$f_{cp}(0)$ exhibits four broad maxima centered at $\theta_0 \simeq 0^\circ,\ 60^\circ,\ 90^\circ,\ \rm{and}\ 120^\circ$.
These four $\theta_0$ all allow formation of both FCC and HCP crystals.
$\theta_0 = 0^\circ,\ 60^\circ,\ \rm{and}\ 120^\circ$ trimers all have $f_{cp}(0) > .5$.
$\theta_0 = 90^\circ$ trimers crystallize far less well:\ their $f_{cp}(0) \simeq 0.1$.
A potential explanation for this difference is suggested by the fact that $\theta_0 = 0^\circ,\ 60^\circ,\ \rm{and}\ 120^\circ$ trimers can tile the 2D triangular lattice, whereas  $\theta_0 = 90^\circ$ trimers cannot.
Specifically, if crystal nuclei preferentially grow by trimers attaching to their triangular ($\{ 1\ 1\ 1 \}$) lattice planes, their growth kinetics will be necessarily be slower for $\theta_0 = 90^\circ$ than for $\theta_0 = 0^\circ,\ 60^\circ,\ \rm{and}\ 120^\circ$.

Panel (a) also illustrates a second key difference between crystal-forming and glass-forming systems.
Most crystal-formers exhibit comparable $f_{\rm FCC}$ and $f_{\rm HCP}$ for $T \lesssim T_s$.
This is not surprising; formation of RHCP-crystalline order is typical for pair interactions ranging from Lennard-Jones to hard-spherical \cite{torquato18}.
What \textit{is} surprising is that formation of FCC (but not HCP) order is strongly suppressed in all glass-forming systems.
The reason for this suppression is clear for $\theta_0 = 33.5^\circ\ \rm{and}\ 70.5^\circ$ -- these trimers can tile the HCP but not the FCC lattice --
and the full $f_{\rm FCC}(T_s)$ dataset suggests that this suppression of FCC relative to HCP ordering extends to $\theta_0$ well away from these special values.
This is a nontrivial result because trimers with any $\theta_0 \leq 120^\circ$ ($\theta_0 \leq 60^\circ$) can form close-packed bilayers (trilayers) \cite{griffith19}.

The $\theta_0$-dependence of the TCC cluster populations at these three characteristic temperatures reinforces the above observations. 
Panel (b) shows three clear maxima [minima] of  $f_{6A}(0)$ [$f_{6Z}(0)$] at $\theta_0 \simeq 0^\circ,\ 60^\circ,\ \rm{and}\ 120^\circ$.
These match the trends in $f_{cp}$ shown in panel (a) -- sensibly so, because small crystalline clusters like 6A support close-packing while small amorphous clusters like 6Z suppress it  \cite{royall15,taffs10,taffs13,malins13a,malins13b}. 
The higher-$T$ results for $f_{6A}$ and $f_{6Z}$ are less conclusive; crystal formers show greater increases in $f_{6A}$ as $T$ approaches $T_s$ from above, but trends in $f_{6Z}$ are not notably different for crystal- vs.\ glass-formers.

Clearer predictive distinctions emerge at the nine-particle level [panel (c)].
For $60^\circ \lesssim \theta_0 \lesssim 100^\circ$, glass-formers have notably higher $f_{9B}$ at $T = T_s + .025\epsilon/k_B$, and the largest $f_{9B}$ for $T \geq T_s$ occur in $\theta_0 = 70.5^\circ$ systems.
Although partial-icosahedral structures like $9B$ are well-known to suppress crystallization in atomic/colloidal systems \cite{frank52,royall15,taffs10,taffs13,malins13a,malins13b}, few previous studies of molecular liquids have explored how these structures' formation propensity varies with the constituent molecules' shape, and this result would have been difficult to predict at the single-trimer level, especially since $\theta_0 = 70.5^\circ$ trimers can close-pack.
Trends in crystalline-cluster populations (e.g. 9K) are less clear, apart from their notably lower $f_{\rm X}(T \gtrsim T_s)$  for $\theta_0 \lesssim 20^\circ$, which is  is consistent with these systems higher $T_s$ and lower $\phi(T_s)$ (Fig.\ \ref{fig:phiTs}).

\begin{figure}[h]
\includegraphics[width=3.2in]{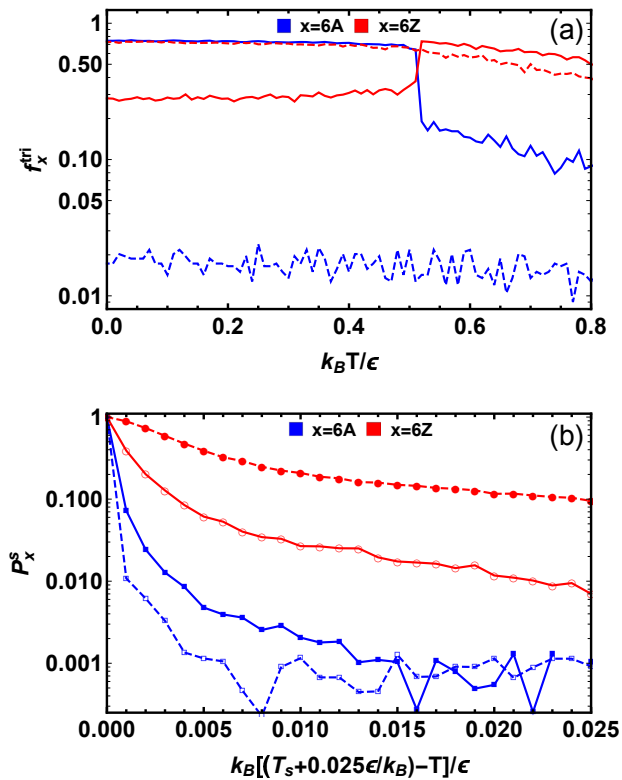}
\caption{Comparison of trimer-level ordering and cluster stability in the best crystal-former (solid curves) vs.\ the best glass-former (dashed curves)..  Panel (a): $f_x^{tri}$.  Panel (b): $P^s_x$.}
\label{fig:fxtri}
\end{figure}

Thus far our cluster-level analyses have treated all monomers equally and ignored dynamics.
This approach is incomplete because we are considering molecular (i.e. trimeric) liquids, and because cluster-level dynamics are critical in determining glass-formability \cite{malins13a,malins13b}.
Figure \ref{fig:fxtri} illustrates the additional insights that can be gained at the six-particle level.
In panel (a), $f_{\rm X}^{tri}$ is the fraction of trimers that lie entirely within a cluster of type X.
Differences between the best crystal-former and best glass-former are far more dramatic than those illustrated in Fig.\ \ref{fig:TCC1}(b).
Although many monomers in the $\theta_0 = 70.5^\circ$ liquid lie within 6A clusters, very few trimers do.
In the $\theta_0 = 120^\circ$ liquid, however, many trimers lie within 6A clusters even at $T = 3T_s/2$.
Thermal diffusive motion of these trimers should make the 6A cluster lifetime much shorter in the $\theta_0 = 70.5^\circ$ liquid than it is in the $\theta_0 = 120^\circ$ liquid.
Panel (b) confirms this hypothesis:\ $P_{\rm X}^s(T)$ is the probability that a given cluster of type $X$ that is present at $T_s + .025\epsilon/k_B$ is still present when the temperature has dropped to $T$.
The data show that 6A (6Z) clusters are far more stable in $\theta_0 = 120^\circ$  ($\theta_0 = 70.5^\circ$) liquids in this temperature range.
Note that these dramatic differences in cluster stability are present despite the fact that the two systems' $T_s$ are almost identical.
While one might argue that they should have been expected because $70.5^\circ$ is very close to the ideal fivefold-symmetry-promoting bond angle ($72^\circ$), we emphasize that they occur despite the fact that  $\theta_0 = 70.5^\circ$ trimers can close-pack \cite{griffith19}.

\section{DIscussion and Conclusions}
\label{sec:conclude}

We examined how thermal solidification of bent-core trimers is influenced by their bond angle $\theta_0$. 
Their solidification temperature $T_s(\theta_0)$ is relatively constant for $\theta \gtrsim 20^\circ$, but increases rapidly with decreasing $\theta_0 \lesssim 20^\circ$ owing to the reduction in trimers' \textit{effective} configurational freedom as they straighten.
This decreasing freedom also produces a sharp decrease in athermal bent-core trimers' $\phi_J(\theta_0)$ as $\theta_0 \rightarrow 0^\circ$ \cite{griffith19}.
We therefore conclude that these systems' thermal-solidification (not just glass-formation) and jamming transitions are intimately connected, in a manner consistent with the ideas of Liu and Nagel \cite{liu98}.
Similar connections have recently been demonstrated for polymers \cite{hoy17,plaza17}.

On the other hand, our results also illustrate a fascinating \textit{contrast} between thermal and athermal solidification.
In athermal 3D systems, the competition between FCC and HCP ordering suppresses crystallization because it produces random-close-packed amorphous order \cite{lubachevsky91,torquato00}.
In thermal trimer solidification, however, it seems that the \textit{absence} of this competition suppresses crystallization.
It is reasonable to suppose that HCP crystallites have a higher nucleation barrier ($\Delta G$) than those of mixed FCC/HCP order owing to their lower entropy. 
A higher $\Delta G$ for trimers that can form only HCP crystals would make them less likely to crystallize  (under cooling at a fixed rate $\dot{T}$) than trimers that can form both FCC and HCP order.
This is indeed what we observe, and is consistent both with Pedersen \textit{et.\ al.}'s finding that Lewis-Wahnstrom-like trimers glass-form more readily than monomeric Lennard-Jones systems primarily because their crystal nucleation barrier is higher \cite{pedersen11,pedersen11b, pedersen19} and with Russo and Tanaka's conclusion that $\beta\Delta G$ is the key quantity controlling glass-formability \cite{russo18}.

We found that bent-core trimers exhibit a wide range of solidification behaviors; some are very good crystal-formers while others are very good glass-formers.
By examining all six of the $\theta_0$ that are commensurable with 3D close-packing, we showed that these differences do not arise from geometric frustration.
Instead, our CCE and TCC analyses showed that they arise from $\theta_0$-dependent differences in trimer liquids' tendencies to form local cluster-level structure that is stable and inhibits crystallization.
Factors promoting greater stability of 6Z and other amorphous clusters (6A and other ordered clusters) unambiguously promote glass-formation (crystallization).
The difference between the present study and previous studies with comparable findings \cite{karayiannis11,karayiannis12,karayiannis13,royall15,taffs10,taffs13,malins13a,malins13b} is that in trimer liquids the relative strength of these factors is determined by the quenched 2- and 3-body constraints (i.e.\ by $R$ and $\theta_0$) rather than by details of the pairwise interactions or preparation protocol.

One of the principal goals of soft materials science is developing materials with tunable solid morphology.
This can be done either by trial-and-error or by developing theories that predict the optimal molecular structure for obtaining a desired morphology and then synthesizing molecules possessing this structure.
The latter approach, now commonly known as ``molecular engineering'', has attracted great interest in recent years and has been applied to systems ranging from photonic crystals \cite{lustig17} to hydrogels \cite{jungst15}.
The results presented in this paper should be a useful contribution to this effort because controlling bond angles in isomeric and near-isomeric  small molecules is a molecular-engineering strategy that is often employed by experimentalists \cite{ping11,powell14,liu17,teerakapibal18}.

This material is based upon work supported by the National Science Foundation under Grant DMR-1555242.
We thank Austin D.\ Griffith for helpful discussions.


\begin{thebibliography}{42}
\bibitem [{oth()}]{otherfrust}%
  \BibitemOpen
  \href@noop {} {}\bibinfo {note} {The other source of frustration is that for
  Lennard-Jones interactions the equilibrium nearest-neighbor distance in a
  zero-pressure zero-$T$ FCC or HCP lattice is $\sim .97\sigma$
  \cite{schwerdtfeger06} rather than $\sigma$ owing to $U_{LJ}(r)$'s long-range
  attractive tail. The effects of this discrepancy are negligible at $T \simeq
  T_s$ because the width of the first peak in the pair correlation function
  $g(r)$ is considerably larger than $.03\sigma$.}\BibitemShut {Stop}%
\bibitem [{gtc()}]{gtcp}%
  \BibitemOpen
  \href@noop {} {}\bibinfo {note} {Systems with Lennard-Jones pair interactions
  can reach $\phi$ values above the hard-sphere close-packing density
  ($\phi_{cp} = \pi/\sqrt{18} \simeq .7405$) because the long-range attractive
  tail of $U_{LJ}(r)$ makes the equilibrium nearest-neighbor distance $r_{nn} <
  \sigma$ in close-packed crystals \cite{schwerdtfeger06}.}\BibitemShut {Stop}%
\bibitem [{foo()}]{footmonsolid}%
  \BibitemOpen
  \href@noop {} {}\bibinfo {note} {Monomers have $T_s = .446\epsilon/k_B$,
  $\phi(0) = .791$, $\phi(T_s) = .697$, and $\phi(T_s + .025\epsilon/k_B) =
  .663$ \cite{gtcp}.}\BibitemShut {Stop}%
\bibitem [{lwc()}]{lwcaveat}%
  \BibitemOpen
  \href@noop {} {}\bibinfo {note} {Direct comparison of our results to those
  for Lewis-Wahnstrom-like models is not straightforward because they have $R =
  2^{1/6}$.}\BibitemShut {Stop}%
\end{thebibliography}

%

\end{document}